# Tracking the Coupling of Single Emitters to Plasmonic Nanoantennas with Single-Molecule Super-Resolution Imaging

Nathan Kimmitt and Esther A. Wertz∗

*Department of Physics, Applied Physics, and Astronomy, Rensselaer Polytechnic Institute Troy, NY 12180, USA).*

E-mail: wertze@rpi.edu

ABSTRACT. Metal nanoantennas enable the manipulation of light emission and detection at the single photon level by confining light into very small volumes. Emitters coupled to these plasmonic structures are thus ideal candidates for usage in quantum information technology. However, controllably and reproducibly placing quantum emitters into nanocavities has been challenging due to the sizes of the systems involved. Here, we investigate the trapping dynamics of single emitters into nanocavities via optical gradient forces by using single-molecule super-resolution imaging and tracking. We show that molecules trapped in the nanogaps of bowtie antennas have increased photostability and track lengths compared to molecules farther away from the structure. For resonant antennas, these effects are magnified compared to the off-resonant case due to stronger optical trapping effects. These results open the way to using plasmonic optical trapping to reliably couple single emitters to single nanoantennas, enabling further studies of these coupled systems.

**Introduction**

The advent of quantum computing has garnered interest in the design and generation of single-photon devices.[1] Indeed, photons are promising as information carriers due to their inherently long decoherence length and speed of associated data transfer. However, photons typically interact very weakly with matter and thus their manipulation usually require high powers and nonlinear media. One way to strengthen these interactions is to use metallic nanoparticles to enhance electromagnetic fields in their vicinity.[2,3] These particles have been shown to concentrate light into nanometer scale volumes and could be used to manipulate photons with low powers in an integratable system.[4] Therefore, understanding the detailed mechanisms of light-matter interactions in these structures is crucial to the development of quantum applications.

Unlike free-space photons, plasmons can be confined below the diffraction limit,[5,6] and thus have the potential to strongly manipulate quantum emitters such as dye molecules or quantum dots. However, they also exhibit considerable ohmic losses, which tend to weaken the plasmon/emitter coupling. Nevertheless, many groups have studied plasmon-exciton coupling and demonstrated the potential of the system in both the weak and strong coupling regimes. In the weak coupling regime, plasmons can strongly enhance emitters' intensities and radiative rates[7-9] and show potential for the development of ultrafast single- photon sources.[10,11] In the strong coupling regime, the exchange of energy between plasmons and excitons becomes reversible, giving rise to hybrid light-matter states,[12,13] which have been theoretically shown to permit novel quantum effects such as single-photon blockade.[14,15] While both of these regimes have been extensively investigated, none of the techniques that were used had the capability to accurately and reproducibly control the position of single emitters in real time. The development of plasmonic optical tweezers[16] allows for the trapping and positional control of nanometer sized objects.[17] However, visualizing the trapping dynamics and where the molecule is positioned remains a challenge due to the size of the systems involved.

In the last decade, single-molecule imaging has developed into a widespread tool to study the effect of plasmonic nanoantennas on the emission of nearby fluorophores.[18–25] One such technique uses freely diffusing molecules as probes by introducing them in a low concentration solution onto the nanoantenna surface.[22–25] These molecules sporadically adsorb to the sample surface at random locations, where their diffusion slows down enough for them to be imaged as a point-spread function (PSF) in the far-field. The total amount of collected photons in each single molecule emission is typically limited by the molecule becoming photobleached. Then, the PSF is fit to a 2D Gaussian to extract the position of the emission and the localization precision depends on the number of photons collected.[26-28] The fluorescence from individual emitters is imaged onto an EMCCD camera over thousands of frames to uniformly sample the surface of the nanoantenna. Here, we demonstrate how plasmonic optical trapping can be used to couple single emitters to nanoantennas and we study the dynamics of the emitter-antenna interactions with single-molecule tracking. By tracking individual emitters over time, we gain new insight into the spatially dependent coupling and show that plasmonic optical trapping can be used to couple emitters to nanoantennas reproducibly for tens of seconds.

## Results

As a model system to study the trapping dynamics of single molecules, we use bowtie nanoantennas (BNAs) that have been shown to give rise to strong field enhancements (Figure 1b).[7,29] Our structures consist of two identically sized gold equilateral triangles, 70 or 100 nanometers in side length, spaced by gap sizes ranging from 10 nm to 30 nm; an SEM image of a typical 70 nm BNA is shown in Figure 1a. Depending on the BNA side-length, the localized surface plasmon resonance (LSPR) can be tuned by over 100 nm from ~ 660 nm to ~ 760 nm (Figure 1c). In this manner, the LSPR for the 70 nm bowtie (Figure 1c, solid red line) – or resonant antenna – has good overlap with the absorption and emission of Cy5 dye (Figure 1c, dashed and solid grey lines, respectively), whereas the LSPR for the 100 nm bowtie (Figure 1c, solid blue line) – or non-resonant antenna – is redshifted from the dye by approximately 100 nm.

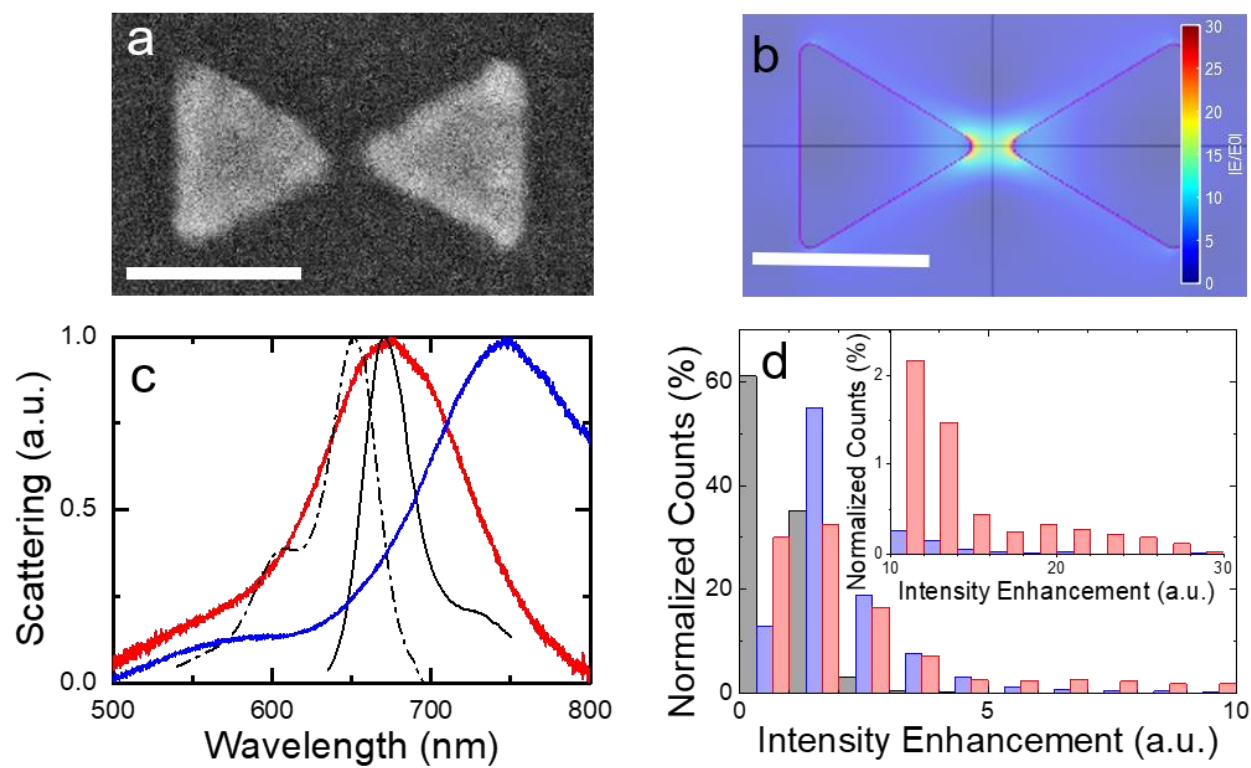

Figure 1: Plasmon enhanced fluorescence with bowtie nanoantennas (BNA). **(a)** Scanning electron microscopy image of a measured 100 nm side length bowtie with a < 10 nm nanogap. **(b)** Finite-element method simulation of electric field enhancement for a nanobowtie of side length 100 nm and 10 nm nanogap. Scale bars: 100 nm. **(c)** Measured darkfield spectra of 70 nm (red line) and 100 nm (blue line) side-length BNA. The Cy5 absorption (dashed black line) and emission (solid black line) spectra are also indicated on the figure. **(d)** Distribution of fluorescence intensities in the presence of resonant (red bar) and non-resonant (blue bar) antennas. The distribution of intensities in the reference case (absence of antenna) is also indicated for comparison, with black bars. Inset shows higher fluorescence enhancement values.

Figure 1d shows a histogram of the fluorescence intensities of single molecules taken in the presence of the resonant and non-resonant BNAs (red and blue bars, respectively) and in the absence of BNA (black line). As expected, in the reference case (Fig 1d black), the distribution of intensities is narrow and does not show any enhancements. In the non-resonant antenna case (Fig 1d blue), the fluorescence enhancement reaches values up to 20 times, and the distribution is much broader. For the resonant antenna (Fig 1d red), we measure enhancements on the order of 30 times, with approximately 6 percent of emitters above ten times enhancement. These are in agreement with previous studies,[19–21,33] where the emission of molecules in the vicinity of an antenna is greatly enhanced compared to the reference case; the enhancement strength depends on the overlap between the LSPR of the antenna and the absorption / emission of the molecule.[22,30]

We performed single-molecule fluorescence measurements of Cy5 in solution interacting with the BNAs based on the technique previously described elsewhere[23] (Methods). Briefly, single molecules adsorbing on the sample surface radiate into the far-field, appearing as a PSF on the camera. We fit this PSF to a 2D-Gaussian and extract the fluorophores' emission positions over thousands of frames. Figure 2 shows the single-molecule localization maps obtained from these experiments. In the reference case (Figure 2a), we see no fluorescence enhancement and a uniform distribution of molecules adsorbing on the surface. In the presence of an antenna (Fig 2b,c), we observe fluorescence enhancement and the emission of the most enhanced molecules are located in the nanogap, at the photonic center of mass. Far from the antenna, we see no coupling to the nanoantenna: those molecules' emission properties are consistent with the reference case.

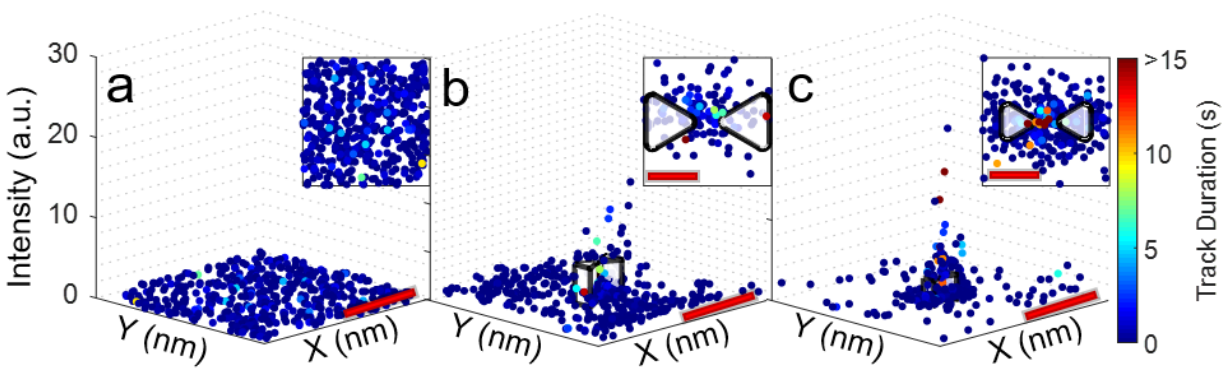

Figure 2: Single-molecule localization maps. Each dot represents the emission position of a single dye molecule. The x- and y-coordinates correspond to the positions with respect to the center of the nearest BNA and the z-axis gives the intensity enhancement of the fit molecule. The color corresponds to the on-time of the single molecule. The data collected over 10000 40 ms frames and about multiple BNAs (typically around 50) are collapsed onto a single unit cell centered around the center of the nanogap. **(a)** Cy5 on a blank ITO coverslip, **(b)** Cy5 near non-resonant antennas, and **(c)** Cy5 near resonant antennas. Scale bar: 500 nm. The gray triangles in the insets for **(b, c)** indicate the location and size of the BNAs. Inset scale bar: 100 nm.

For molecules in the near-field of the antenna, the combined effect of the available local density of states (LDOS) and electromagnetic field enhancement leads to enhanced absorption as well as emission, resulting in more photons detected in the far-field.[22] We used the COMSOL Multiphysics package (Methods) to get an expected fluorescence enhancement value for the emission of the fluorophores. The results agree well with our experimental data for both the resonant and non-resonant antenna, demonstrating that the electric field enhancement, which is confined to the nanogap, is the driving factor in fluorescence enhancement. Figure 2 further shows that emitters placed in the vicinity of the antenna display increased on-time based off position (i.e. in the nanogap versus outside) as well as antenna size. To further investigate this dependence, we examine fluorescence time traces of individual molecules.

We tracked individual molecules over time using an in-house Matlab code (Methods): after obtaining the positions of emitters over thousands of frames, we form tracks by implementing the Hungarian algorithm to optimize all possible pairings of fit emitters in successive frames. Single-molecule tracks are shown in Figure 3 as functions of position and intensity. In these figures each dot represents the molecule's average position over ten frames in order to highlight its overall movement direction which is obscured by the Brownian motion of the particle from frame to frame. The top row shows the fluorescence intensity's dependence on the BNA long-axis (longitudinal position), while the bottom row shows a top view of the molecule's position within the x-y plane; time is represented by color with earlier positions in blue and later in red.

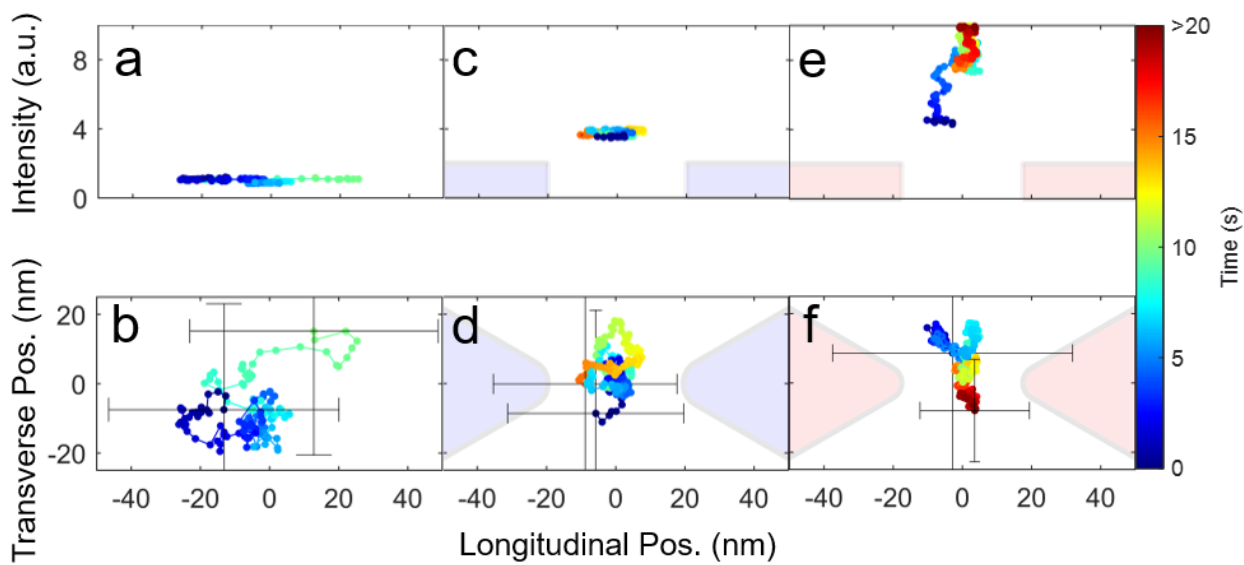

Figure 3: Single-molecule tracks. Points were averaged over 10 frames to highlight long-term trends in the molecule's motion. The top row shows intensity enhancement as a function of longitudinal position while the bottom row shows single molecule trajectories; color represents time. **(a-b)** Reference case; the molecule shows a constant intensity and its motion isn't directional or confined. **(c-d)** Non-resonant case; the molecule's motion shows tighter confinement to the nanogap as well as enhanced intensity and emission duration. **(e-f)** Resonant case; the molecule's intensity is highly dependent on its distance from the nanogap and is tightly confined over large time scales. The blue triangles in **(c,d)** and red triangles in **(e,f)** indicate the BNA and nanogap sizes and positions.

For molecules adsorbing onto the bare ITO coverslip in the absence of BNAs, the single-molecule tracks are typically short with only small intensity variations (Figure 3a) and little confinement (Figure 3b), as expected from the Brownian motion of an uncoupled molecule. In the case of the non-resonant BNA, molecules tracked in the nanogap show a tighter confinement (Figure 3d) and enhanced intensities (Figure 3c). Finally, Figure 3e and 3f show a molecule adsorbed near the resonant BNA nanogap. As the molecule approaches the nanogap, its intensity slowly increases over time, reaching its maximum as the molecule fully enters the cavity (Figure 3e). Once within the nanogap, the molecule remains confined there (Figure 3f) and shows sporadic intensity fluctuations.

To quantitatively understand these trends, we analyzed the time traces of hundreds of molecules in the three regimes described above: in the reference, non-resonant BNA and resonant BNA cases. Representative time traces for these regimes are plotted in Figure 4. Molecules tracked in the reference case, far from any antenna show a typical single-molecule behavior: a stepwise increase in intensity to a constant value, lasting no more than five seconds (Figure 4a and Figure 4d, black bar). When the molecule photobleaches, a stepwise decrease back to the background noise is observed. Figure 4b shows representative time traces for molecules tracked in the nanogap of non-resonant antennas. In this case, significantly larger intensity values are observed along with increased track durations (Figure 4a and Figure 4d, blue bar). The characteristic stepwise increase and decrease is still evident. Figure 4c shows representative time traces for molecules within the nanogap of resonant antennas. Molecules here exhibit sustained (on the order of tens of seconds) fluorescence enhancement (Figure 4d, red bar) with large intensity fluctuations. This is consistent with constrained, but not prohibited motion of the particle. As it moves through areas of high electromagnetic field, the intensity fluctuates. For antenna with higher electric field gradients, we expect higher intensity fluctuations.

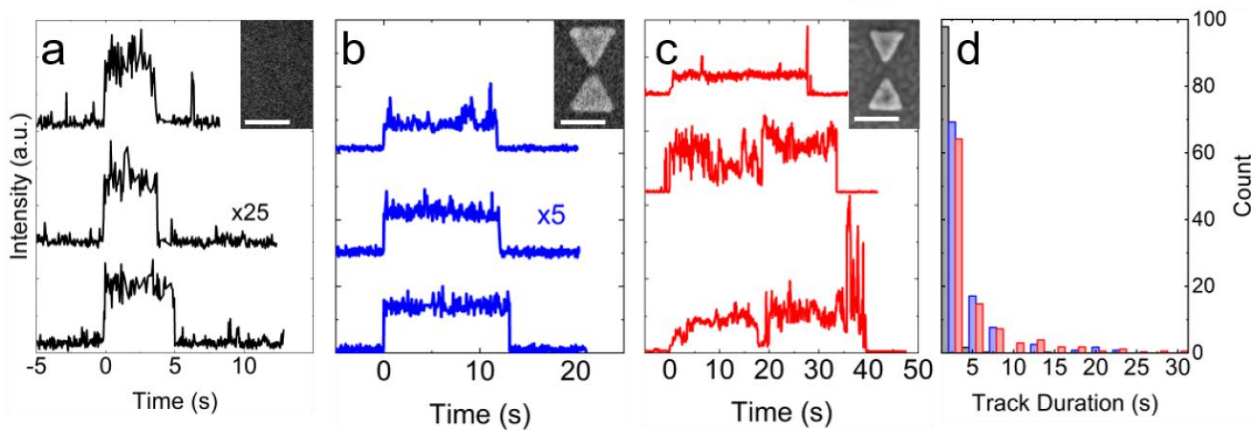

Figure 4: Single-molecule fluorescence time traces. **(a)** Cy5 on a blank ITO coverslip, **(b)** Cy5 near non-resonant antennas, and **(c)** Cy5 near resonant antennas. In the reference case, on-times are typically less than five seconds while molecules within BNA nanogaps – both non-resonant (blue lines) and resonant (red lines) – show increased on-times and enhanced intensities. Insets show representative SEM images environment around which the time traces were measured; scale bars: 100nm. **(d)** Histogram of track durations in the reference case (black bars), non-resonant antenna case (blue bars) and resonant antenna case (red bars).

These time traces show that overall, the fluorophores in the presence of the nanoantennas experience longer emission times than those in the reference case. Indeed, no track exceeds five seconds in the reference case but five second tracks or longer represent over 30 percent of tracks in the presence of BNAs, both resonant and non-resonant. Typical fluorophores can be approximated by a three-level model, where there is a non-zero probability of intersystem crossing into the triplet state.[31] Molecules in the nanogap experience higher radiative and absorption decay rates, but the intersystem crossing rate remains unaltered so that overall, the probability of intersystem crossing decreases and molecules emit for longer times.[32] We see that the on-times differ between the resonant and non-resonant cases by approximately a factor of two, matching the difference in intensity enhancement between the two cases. Both effects are a direct consequence of the increased radiative rate of the emitter when placed in the nanogap.

Concurrently, the BNAs generate enhanced optical gradient forces leading to tighter confinement of the molecular tracks in the nanogaps, which decreases the spread of the tracks.

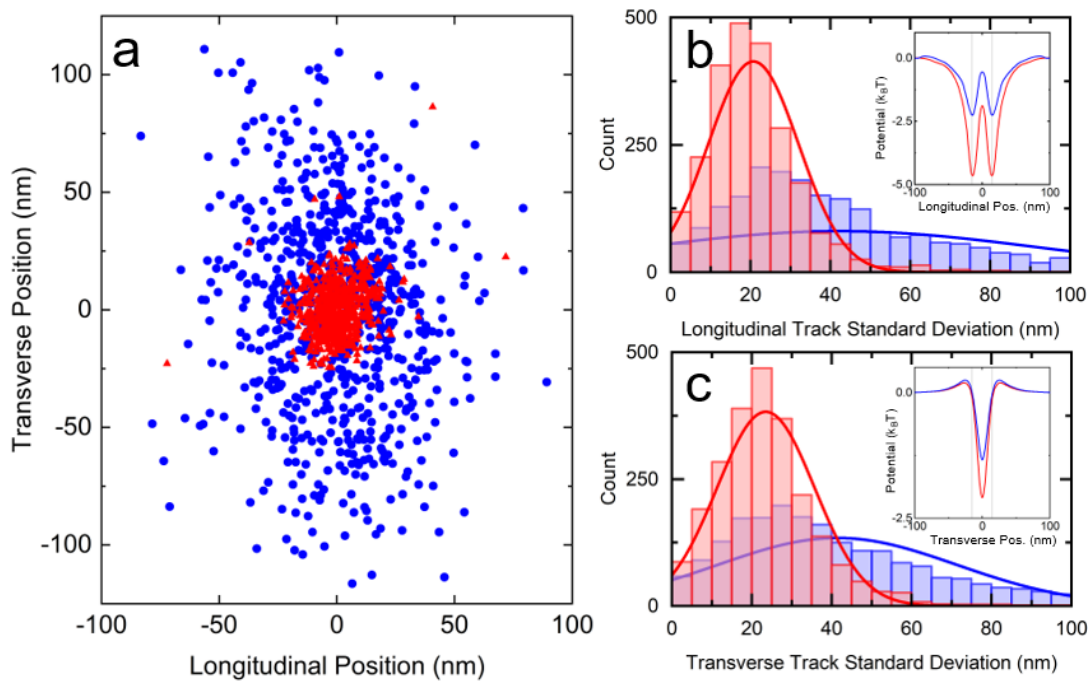

Figure 5: Plasmonic optical trapping. **(a)** Each point represents the position of a tracked molecule for a resonant (red) and non-resonant (blue) antenna. The resonant antenna displays tighter confinement to the nanogap. **(b,c)** Tracked molecules' standard deviation histograms and corresponding normal fits along the longitudinal and transverse directions, respectively. Red: resonant, blue: non-resonant. The confinement is strongest along the longitudinal direction for resonant antennas. Data compiled over 16 different sets of resonant and non-resonant antennas. Insets: Finite element simulations of the trapping potential along the longitudinal **(b)** and transverse **(c)** direction, respectively for non-resonant (blue) and resonant (red) antennas, 10 nm above the structures.

Figure 5a shows a scatterplot of the longitudinal and transverse position of a tracked molecule for the non-resonant (blue) and resonant (red) antenna; Figures 5b and c are histograms of tracked molecules' standard deviations along the longitudinal (Figure 5b), and transverse (Figure 5c) directions. These histograms represent thousands of tracked molecules taken from several sets of resonant and non-resonant antennas; smaller standard deviations represent more confined tracks.

The motion of the molecule tracked in Figure 3f is directed towards the nanogap: as time passes, the molecule moves towards the center of the cavity. This is consistent with the plasmonic optical trapping effect which tends to pull nearby emitters towards the maximum of the electromagnetic field.[16,17,33] Indeed, the very tight focusing of the electromagnetic fields promotes strong optical gradient forces in these BNA structures. In the dipole approximation, the time-averaged optical gradient force scales as the gradient of the square of the electric field. Using finite-element simulations we determined the trapping potential for the antenna system along both axis directions. Figure 5b,c insets show the result along the longitudinal and transverse directions, respectively, 10 nm above the structure for the resonant (red) and non-resonant (blue) case. The trapping potential along this direction is largest at the tips and is on the order of the accepted value of ~ 10 $k_B$T accounting for random fluctuations in the particle's Brownian motion needed for stable trapping.[34] The depth of these potential wells depends not only on distance above the structures, but on antenna size as well. More resonant structures have higher electric field gradients, resulting in deeper potential wells that constrain molecule motion. Emitters in the nanogap experience these optical trapping forces, resulting in strong confinement where the strength of the coupling to the antenna is largest. The increased plasmon enhancement in the nanogapis evidenced by the large fluorescence enhancement values and track durations. Thus, freely diffusing particles may become temporarily confined in the nanogap of the bowtie structures where the coupling is strongest and the strength of this confinement can be tuned by maximizing the electromagnetic field strength.

**Conclusion**

We studied the coupling of single emitters to single antennas using an optical microscope and a simple geometry. By implementing super-resolution imaging techniques, we were able to track individual molecules, gaining new insight into the intrinsic plasmonic optical trapping ability of these systems. We see these effects via tighter confinement of single molecules for resonant antennas versus non-resonant antennas. Molecules placed in the nanogap experience increased absorption and radiative rates, resulting in increased on-times. Plasmonic optical trapping can be utilized to couple emitters to nanoantennas reproducibly for tens of seconds, and to obtain tight confinement (∼ 10 nm) within the structures nanogap.

We obtained our results with an unoptimized and simple geometry. More optimized geometries will have even stronger effects on the confinement and emission properties of single emitters. This demonstrates the feasibility of fabricating components such as single-photon switches and transistors in easily integratable systems. Additionally, our study opens the pathway to probe the strong regime in more optimized geometries.

**Methods**

**Sample fabrication.** A 100 nm thick layer of indium tin oxide was sputtered onto a clean glass coverslip. Positive photoresist (ZEP520A (1:1)) was spun on at 4000 RPM for 60 s (∼150nm thickness) and baked at 180 C for 300 s. To facilitate the electron beam lithography process, an additional conductive layer (e-spacer 300Z) was spun onto the ITO coverslips at 3000 RPM for 60s (∼100nm thickness) and baked at 110 C for 180 s. We performed electron beam lithography using the JEOL-6300FS at 100 kV using 1 nA current and patterned the different-sized BNAs into periodic arrays, with 1 $\mu$m separations in both directions. The samples were developed in hexyl acetate at -20 C for 120 s, followed by an isopropanol rinse for 30 s and a N2 dry. After development, we deposited a 3 nm Ti wetting layer and a 30 nm Au layer onto the surface by electron beam physical vapor deposition. The remaining photoresist was lifted off by soaking in N-methyl-2-pyrrolidone for ∼ 12 h at 90 C, and then sonicating in acetone for 30 s. We have verified by "before and after" darkfield spectroscopy measurements that the antennas remain unaltered over the course of our single-molecule imaging experiments.

**PAINT/Trapping Experimental Setup.** We used Amersham CyDye mono-reaction NHS esters (Cy5) as fluorescent probes in our experiments. Widefield epifluorescence images of single Cy5 molecules were acquired through an Olympus IX-83 inverted microscope. A circularly polarized, 640 nm continuous-wave laser exited the sample through a 100x, 1.4 NA Olympus oil-immersion microscope objective, which was also used to collect the single-molecule fluorescence. A long-pass filter (Semrock FF01-692/40-25) and a dichroic mirror (Semrock FF660-Di02-25x36) eliminated scattered laser light, and the image was projected via a 2x beam expander onto a Princeton ProEM-HS CCD Camera at 1000x EM gain. To improve statistics, the data from multiple nanoantennas was collapsed onto a relative coordinate system.

**Finite-Element Simulations.** Finite element simulations were run using the COMSOL Multiphysics Program. The simulations were performed using a combination of the full-wave and scattered-wave simulation formalism. A uniform plane wave was generated at the upper boundary, and the effects of this wave scattering on the substrate without the presence of the nanoparticles was taken as the background field for the scattering formalism. Perfectly matched layers (PML) were placed at both the top and bottom of the domain to prevent backscattering effects from the boundaries. The absorption cross-section was calculated by integrating the power loss density over the volume of the bowtie. The scattering cross section was calculated by integrating the relative Poynting vector over the surface of the bowtie.

**Molecular Tracking.** Tracking of individual molecules is based off two main assumptions. First, that no two molecules are ever emitting within the same diffraction limited area. Second, that if a molecule shifts a certain physical distance between frames (i.e. more than 100 nm) or is in the dark state for more than 15 frames (0.6s), the track is ended. After the fit positions are found over thousands of frames using our in-house MATLAB code, we use an algorithm to give a unique numerical identifier to individual molecules. The position of these tagged molecules is monitored over time, forming tracks.

AUTHOR INFORMATION

**Corresponding Author**

* E-mail: wertze@rpi.edu

**Author Contributions**

N.K. performed the experiments and finite element simulations, analyzed the data, and wrote the paper. N.K and E.A.W. designed the experiments, discussed the results, and edited the paper.

ACKNOWLEDGMENT

This work was supported by the National Science Foundation CAREER award (DMR1945035). This research used resources of the Center for Functional Nanomaterials, which is a U.S. DOE Office of Science Facility, at Brookhaven National Laboratory under Contract No. DE-SC0012704.

REFERENCES

(1) Brassard, G.; Chuang, I.; Lloyd, S.; Monroe, C. Quantum computing. *Proceedings of the National Academy of Sciences* **1998**, *95*, 11032–11033.

(2) Hugall, J. T.; Singh, A.; van Hulst, N. F. Plasmonic cavity coupling. *ACS Photonics* **2018**, *5*, 43–53.

(3) Marquier, F.; Sauvan, C.; Greffet, J.-J. Revisiting quantum optics with surface plasmons and plasmonic resonators. *ACS Photonics* **2017**, *4*, 2091–2101.

(4) Novotny, L.; Van Hulst, N. Antennas for light. *Nature Photonics* **2011**, *5*, 83–90.

(5) Gramotnev, D. K.; Bozhevolnyi, S. I. Plasmonics beyond the diffraction limit. *Nature Photonics* **2010**, *4*, 83–91.

(6) Schuller, J. A.; Barnard, E. S.; Cai, W.; Jun, Y. C.; White, J. S.; Brongersma, M. L. Plasmonics for extreme light concentration and manipulation. *Nature materials* **2010**, *9*, 193–204.

(7) Kinkhabwala, A.; Yu, Z.; Fan, S.; Avlasevich, Y.; Mü̈llen, K.; Moerner, W. Large single-molecule fluorescence enhancements produced by a bowtie nanoantenna. *Nature Photonics* **2009**, *3*, 654–657.

(8) Akselrod, G. M.; Argyropoulos, C.; Hoang, T. B.; Ciraci, C.; Fang, C.; Huang, J.; Smith, D. R.; Mikkelsen, M. H. Probing the mechanisms of large Purcell enhancement in plasmonic nanoantennas. *Nature Photonics* **2014**, *8*, 835–840.

(9) Baumberg, J. J.; Aizpurua, J.; Mikkelsen, M. H.; Smith, D. R. Extreme nanophotonics from ultrathin metallic gaps. *Nature Materials* **2019**, *18*, 668–678.

(10) Hoang, T. B.; Akselrod, G. M.; Mikkelsen, M. H. Ultrafast room-temperature single photon emission from quantum dots coupled to plasmonic nanocavities. *Nano Letters* **2016**, *16*, 270–275.

(11) Bogdanov, S. I.; Makarova, O. A.; Xu, X.; Martin, Z. O.; Lagutchev, A. S.; Olinde, M.; Shah, D.; Chowdhury, S. N.; Gabidullin, A. R.; Ryzhikov, I. A., et al. Ultrafast quantum photonics enabled by coupling plasmonic nanocavities to strongly radiative antennas. *Optica* **2020**, *7*, 463–469.

(12) Chikkaraddy, R.; De Nijs, B.; Benz, F.; Barrow, S. J.; Scherman, O. A.; Rosta, E.; Demetriadou, A.; Fox, P.; Hess, O.; Baumberg, J. J. Single-molecule strong coupling at room temperature in plasmonic nanocavities. *Nature* **2016**, *535*, 127–130.

(13) Santhosh, K.; Bitton, O.; Chuntonov, L.; Haran, G. Vacuum Rabi splitting in a plasmonic cavity at the single quantum emitter limit. *Nature Communications* **2016**, *7*, 1–5.

(14) Chang, D. E.; Sørensen, A. S.; Demler, E. A.; Lukin, M. D. A single-photon transistor using nanoscale surface plasmons. *Nature Physics* **2007**, *3*, 807–812.

(15) Manjavacas, A.; Nordlander, P.; Garcia de Abajo, F. J. Plasmon blockade in nanostructured graphene. *ACS nano* **2012**, *6*, 1724–1731.

(16) Novotny, L.; Bian, R. X.; Xie, X. S. Theory of nanometric optical tweezers. *Physical Review Letters* **1997**, *79*, 645.

(17) Xu, Z.; Song, W.; Crozier, K. B. Direct particle tracking observation and Brownian dynamics simulations of a single nanoparticle optically trapped by a plasmonic nanoaperture. *ACS Photonics* **2018**, *5*, 2850–2859.

(18) Stranahan, S. M.; Willets, K. A., Super-resolution Optical Imaging of Single-Molecule SERS Hot Spots. *Nano Letters* **2011**, 10, 3777-3784.

(19) Su, L.; Yuan, H.; Lu, G.; Rocha, S.; Orrit, M.; Hofkens, J.; Uji-i, H. Super-resolution localization and defocused fluorescence microscopy on resonantly coupled single-molecule, single-nanorod hybrids. *ACS Nano* **2016**, *10*, 2455–2466.

(20) Raab, M.; Vietz, C.; Stefani, F. D.; Acuna, G. P.; Tinnefeld, P. Shifting molecular localization by plasmonic coupling in a single-molecule mirage. *Nature Communications* **2017**, *8*, 13966.

(21) Singh, A.; de Roque, P. M.; Calbris, G.; Hugall, J. T.; van Hulst, N. F. Nanoscale mapping and control of antenna-coupling strength for bright single photon sources. *Nano Letters* **2018**, *18*, 2538–2544.

(22) Mack, D. L.; Cortes, E.; Giannini, V.; Török, P.; Roschuk, T.; Maier, S. A. Decoupling absorption and emission processes in super-resolution localization of emitters in a plasmonic hotspot. *Nature Communications* **2017**, *8*, 1–10.

(23) Wertz, E.; Isaacoff, B. P.; Flynn, J. D.; Biteen, J. S. Single-molecule super-resolution microscopy reveals how light couples to a plasmonic nanoantenna on the nanometer scale. *Nano Letters* **2015**, *15*, 2662–2670.

(24) Taylor, A.; Verhoef, R.; Beuwer, M.; Wang, Y.; Zijlstra, P. All-optical imaging of gold nanoparticle geometry using super-resolution microscopy. *The Journal of Physical Chemistry C* **2018**, *122*, 2336–2342.

(25) Cang, H.; Labno, A.; Lu, C.; Yin, X.; Liu, M.; Gladden, C.; Liu, Y.; Zhang, X. Probing the electromagnetic field of a 15-nanometre hotspot by single molecule imaging. *Nature* **2011**, *469*, 385–388.

(26) Moerner, W. E. Single-molecule mountains yield nanoscale cell images. *Nature Methods* **2006**, *3*, 781-782.

(27) Thompson, R. E.; Larson, D. R.; Webb, W. W. Precise nanometer localization analysis for individual fluorescent probes. *Biophysical journal* **2002**, *82*, 2775–2783.

(28) Mortensen, K. I.; Churchman, L. S.; Spudich, J. A.; Flyvbjerg, H. Optimized localization analysis for single-molecule tracking and super-resolution microscopy. *Nature methods* **2010**, *7*, 377–381.

(29) Ko, K. D.; Kumar, A.; Fung, K. H.; Ambekar, R.; Liu, G. L.; Fang, N. X.; Toussaint Jr, K. C. Nonlinear optical response from arrays of Au bowtie nanoantennas. *Nano Letters* **2011**, *11*, 61–65.

(30) Wertz, E. A.; Isaacoff, B. P.; Biteen, J. S. Wavelength-dependent super-resolution images of dye molecules coupled to plasmonic nanotriangles. ACS Photonics 2016, 3, 1733-1740.

(31) Eggeling, C.; Widengren, J.; Rigler, R.; Seidel, C. Photobleaching of fluorescent dyes under conditions used for single-molecule detection: Evidence of two-step photolysis. *Analytical Chemistry* **1998**, *70*, 2651–2659.

(32) Novotny, L.; Hecht, B. *Principles of nano-optics*; Cambridge university press, 2012.

(33) Roxworthy, B. J.; Ko, K. D.; Kumar, A.; Fung, K. H.; Chow, E. K.; Liu, G. L.; Fang, N. X.; Toussaint Jr, K. C. Application of plasmonic bowtie nanoantenna arrays for optical trapping, stacking, and sorting. *Nano Letters* **2012**, *12*, 796–801.

(34) Huang, J.-S.; Yang, Y.-T. Origin and future of plasmonic optical tweezers. *Nanomaterials* **2015**, *5*, 1048–1065.